\documentclass[pdflatex,sn-mathphys-num]{sn-jnl}% Math and Physical Sciences Numbered Reference Style 
\usepackage{graphicx}%
\usepackage{multirow}%
\usepackage{amsmath,amssymb,amsfonts}%
\usepackage{amsthm}%
\usepackage{mathrsfs}%
\usepackage[title]{appendix}%
\usepackage{xcolor}%
\usepackage{textcomp}%
\usepackage{manyfoot}%
\usepackage{booktabs}%
\usepackage{algorithm}%
\usepackage{algorithmicx}%
\usepackage{algpseudocode}%
\usepackage{listings}%
\theoremstyle{thmstyleone}%
\theoremstyle{thmstyletwo}%

\theoremstyle{thmstylethree}%

\usepackage{lineno}

\begin{document}
\title[Article Title]{Towards a Unified Benchmark and Framework for Deep Learning-Based Prediction of Nuclear Magnetic Resonance Chemical Shifts}

%%=============================================================%%
%% GivenName	-> \fnm{Joergen W.}
%% Particle	-> \spfx{van der} -> surname prefix
%% FamilyName	-> \sur{Ploeg}
%% Suffix	-> \sfx{IV}
%% \author*[1,2]{\fnm{Joergen W.} \spfx{van der} \sur{Ploeg} 
%%  \sfx{IV}}\email{iauthor@gmail.com}
%%=============================================================%%
% \linenumbers

\author[1,2]{\fnm{Fanjie} \sur{Xu}}\email{xufanjie@stu.xmu.edu.cn}

\author[2,3]{\fnm{Wentao} \sur{Guo}}\email{wtguo@ucdavis.edu}

\author[1]{\fnm{Feng} \sur{Wang}}\email{fengwang@xmu.edu.cn}

\author[2]{\fnm{Lin} \sur{Yao}}\email{yaol@dp.tech}

\author[2]{\fnm{Hongshuai} \sur{Wang}}\email{wanghongshuai@dp.tech}

\author*[4,5]{\fnm{Fujie} \sur{Tang}}\email{tangfujie@xmu.edu.cn}

\author*[2]{\fnm{Zhifeng} \sur{Gao}}\email{gaozf@dp.tech}

\author[2,6]{\fnm{Linfeng} \sur{Zhang}}\email{zhanglf@dp.tech}

\author[6,7,8]{\fnm{Weinan} \sur{E}}\email{weinan@math.pku.edu.cn}

\author[1,5]{\fnm{Zhong-Qun} \sur{Tian}}\email{zqtian@xmu.edu.cn}

\author*[1,5,9]{\fnm{Jun} \sur{Cheng}}\email{chengjun@xmu.edu.cn}

\affil*[1]{\orgdiv{State Key Laboratory of Physical Chemistry of Solid Surface}, \orgdiv{iChEM}, \orgdiv{College of Chemistry and Chemical Engineering}, \orgname{Xiamen University}, \orgaddress{ \city{Xiamen}, \postcode{361005}, \country{China}}}

\affil[2]{\orgname{DP Technology}, \orgaddress{\city{Beijing}, \postcode{100080}, \country{China}}}

\affil[3]{\orgdiv{Department of Chemistry}, \orgname{University of California}, \orgaddress{\city{Davis}, \postcode{95616}, \country{USA}}}

\affil[4]{\orgdiv{Pen-Tung Sah Institute of Micro-Nano Science and Technology}, \orgname{Xiamen University}, \orgaddress{ \city{Xiamen}, \postcode{361005}, \country{China}}}

\affil[5]{\orgdiv{Laboratory of AI for Electrochemistry (AI4EC)}, \orgname{Tan Kah Kee Innovation Laboratory (IKKEM)}, \orgaddress{ \city{Xiamen}, \postcode{361005}, \country{China}}}

\affil[6]{\orgname{AI for Science Institute}, \orgaddress{\city{Beijing}, \postcode{100080}, \country{China}}}

\affil[7]{\orgdiv{Center for Machine Learning Research}, \orgname{Peking University}, \orgaddress{ \city{Beijing}, \postcode{100871}, \country{China}}}

\affil[8]{\orgdiv{School of Mathematical Sciences}, \orgname{Peking University}, \orgaddress{ \city{Beijing}, \postcode{100871}, \country{China}}}

\affil[9]{\orgdiv{Institute of Artificial Intelligence}, \orgname{Xiamen University}, \orgaddress{ \city{Xiamen}, \postcode{361005}, \country{China}}}

%%==================================%%
%% Sample for unstructured abstract %%
%%==================================%%

\abstract{The study of structure-spectrum relationships is essential for spectral interpretation, impacting structural elucidation and material design. Predicting spectra from molecular structures is challenging due to their complex relationships. Herein, we introduce NMRNet, a deep learning framework using the SE(3) Transformer for atomic environment modeling, following a pre-training and fine-tuning paradigm. To support the evaluation of NMR chemical shift prediction models, we have established a comprehensive benchmark based on previous research and databases, covering diverse chemical systems. Applying NMRNet to these benchmark datasets, we achieve state-of-the-art performance in both liquid-state and solid-state NMR datasets, demonstrating its robustness and practical utility in real-world scenarios. This marks the first integration of solid and liquid state NMR within a unified model architecture, highlighting the need for domain-specific handling of different atomic environments. Our work sets a new standard for NMR prediction, advancing deep learning applications in analytical and structural chemistry.}


%%\pacs[JEL Classification]{D8, H51}

%%\pacs[MSC Classification]{35A01, 65L10, 65L12, 65L20, 65L70}

\maketitle

\section*{Introduction}\label{sec:introduction}

Spectroscopy plays a crucial role for elucidating molecular structures and dynamics\cite{ xue2023advances, lu2024deep}. Nuclear Magnetic Resonance (NMR) is a powerful technique widely used in chemistry, biology, and materials science to reveal the local chemical environment\cite{alderson2021nmr, atwi2022automated}. Accurate prediction of NMR chemical shifts can significantly aid in the assignment of spectrum signals and the interpretation of NMR spectra\cite{chen2020review, merrill2023nmr, hu2023machine}, further enhances structure revision and configuration determination\cite{smith2010assigning, tsai2022ml, novitskiy2022du8ml, novitskiy2022brief}. Traditional methods for NMR chemical shift prediction, while valuable, often encounter limitations in balancing accuracy and efficiency, especially when dealing with intricate molecular architectures\cite{jonas2022prediction, cortes2023machine}. Recent advancements in deep learning offer promising avenues for enhancing NMR chemical shift predictions\cite{gerrard2020impression, yang2021predicting, guan2021real}.

Current deep learning approaches for chemical shift prediction address two main topics: liquid-state and solid-state NMR. For liquid-state NMR, two of the largest publicly available database are nmrshiftdb2\cite{kuhn2012chemical, kuhn2015facilitating} and QM9-NMR\cite{gupta2021revving}. After applying restrictions on the number of atoms and types of elements in the molecules within nmrshiftdb2, Kuhn et al.\cite{jonas2019rapid} extracted the remaining molecules and their corresponding chemical shifts to create a derivative dataset (we refer to it as nmrshiftdb2-2018). They utilized a graph convolutional network (GCN), achieving better results compared to traditional HOSE code methods\cite{bremser1978hose}. More recently, DetaNet\cite{zou2023deep}, based on the E(3)-equivariant Message Passing Neural Network (MPNN), has demonstrated significant improvements on the QM9-NMR dataset compared to previous work using FCHL\cite{gupta2021revving, christensen2020fchl}. Liquid-state NMR models often neglect intermolecular and solute-solvent interactions, focusing on individual molecules. 

For solid-state NMR, the lack of well-organized and open-access experimental datasets\cite{paruzzo2018chemical} necessitates reliance on computational datasets. Unlike in liquid-state NMR, extracting each atom's local environment in solid-state NMR must be conducted under periodic boundary conditions (PBC). In 2018, Paruzzo et al.\cite{paruzzo2018chemical, cordova2022machine} compiled structures from the Cambridge Structural Database (CSD)\cite{groom2016cambridge} and employed DFT to obtain chemical shifts. They trained the ShiftML model using smooth overlap of atomic positions (SOAP)\cite{bartok2013representing} descriptors. Later, Cheng et al. proposed the NN-NMR model for predicting the NMR of battery materials\cite{lin2021unravelling, lin2022combining, lin2022machine}, also using SOAP descriptors, with configuration sampling by DPMD\cite{zhang2018deep} and calculations by DFT. Given the distinct characteristics of liquid-state and solid-state NMR, separate frameworks are currently required to accurately predict NMR chemical shifts. However, an ideal approach would be to develop a unified framework capable of effectively processing data from both states. 

In recent years, the pre-training and fine-tuning paradigm has achieved significant success in natural language processing\cite{devlin2018bert, liu2019roberta} and computer vision\cite{dosovitskiy2020image, kirillov2023segany}, particularly in large language models\cite{achiam2023gpt, touvron2023llama}. This approach has also begun to advance research and applications in fundamental science\cite{jumper2021highly, lin2022language, zhou2023uni, zhang2023dpa}. By pre-training on large-scale, unlabeled biological and chemical datasets, models can acquire extensive knowledge, which can then be fine-tuned for specific tasks such as predicting biochemical properties and structural generation. Inspired by this success, we adapted this paradigm to the NMR prediction task.

In this study, we present NMRNet, a unified framework through a pre-training and fine-tuning paradigm, designed for general atomic environment modeling based on a shared SE(3) Transformer architecture. By collecting extensive crystal structure data for pre-training, we established an efficient representation of the local environment of crystals, enabling direct application to various downstream tasks related to crystalline materials. Additionally, we assembled a comprehensive benchmark for evaluating NMR chemical shift prediction models by gathering a diverse dataset of chemical systems widely used in previous research. This included the creation of a standardized dataset, nmrshiftdb2-2024, derived from extensive cleaning and validation of the existing NMR experimental database, nmrshiftdb2, serving as a foundational resource for future advancements in NMR prediction models. Our results demonstrate that NMRNet achieves state-of-the-art predictive performance in both liquid-state and solid-state NMR, exhibiting robust generalization capabilities even in complex systems. Furthermore, we highlight the potential of NMRNet for spectral interpretation and structural analysis, underscoring its practical utility in real-world scenarios. This work brings new perspectives to the field of spectroscopy and structural analysis, significantly contributing to future innovations in material design and chemical research.

\begin{figure}
    \centering
    \includegraphics[width=0.96\linewidth]{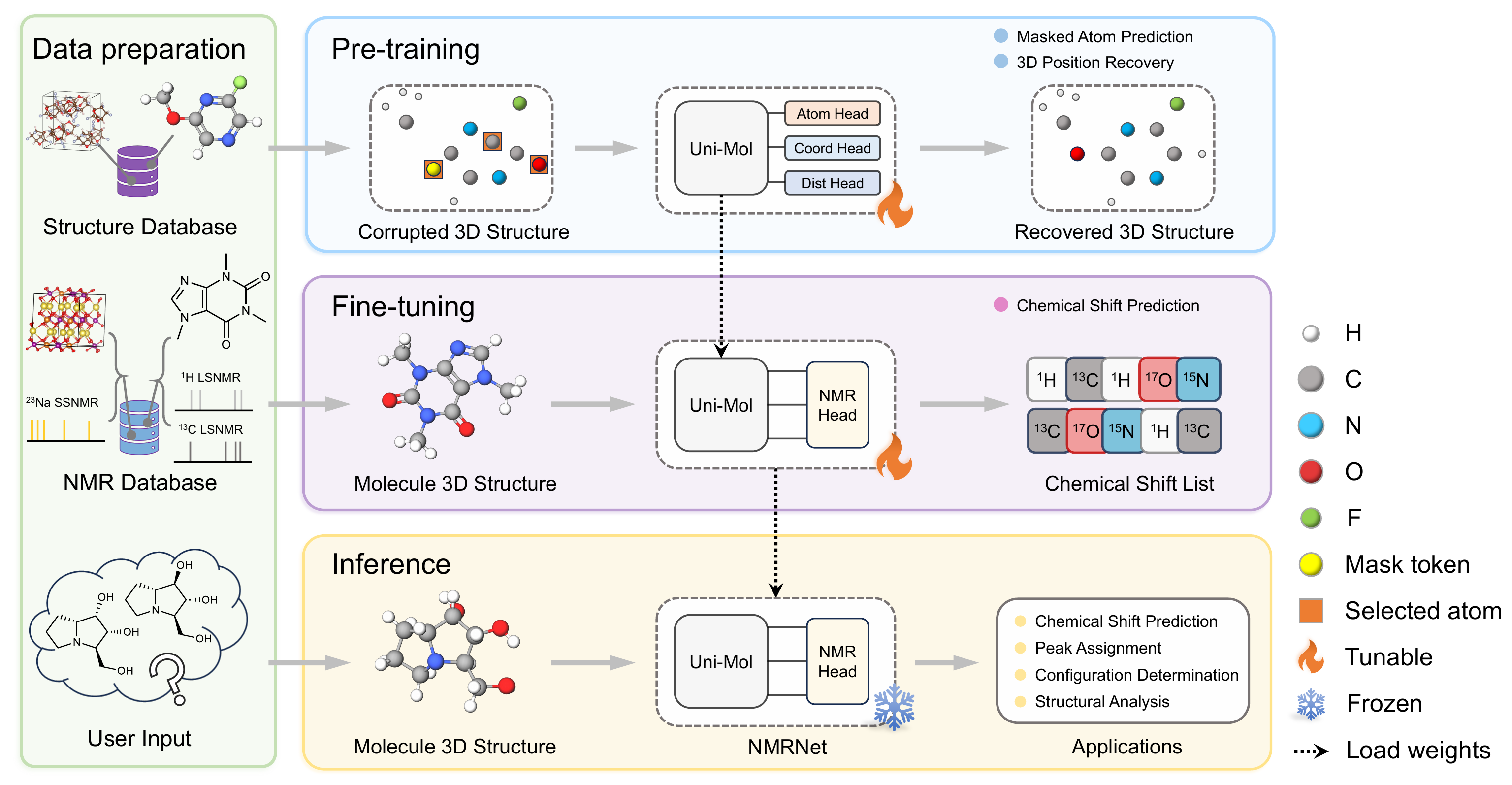}
    \caption{\textbf{Schematic diagram of NMRNet framework.} Left: Data preparation, providing structure and NMR data. Right top: Pre-training, using pure structural information for self-supervised tasks, including masked atom prediction and 3D position recovery. Right middle: Fine-tuning, for supervised NMR chemical shift prediction. Right bottom: Inference, where the fine-tuned NMRNet model parameters are frozen and applied to various tasks.}
    \label{fig:framework}
\end{figure}

\section*{Results and discussion}
\label{sec:results}

\subsection*{Overview of NMRNet framework}
The NMRNet framework comprises four modules: data preparation, pre-training, fine-tuning, and inference. The overall workflow is illustrated in Fig. \ref{fig:framework}. The molecular domain is introduced by Uni-Mol\cite{zhou2023uni}, a representation framework with a SE(3) Transformer architecture that has been extensively used for molecular downstream tasks\cite{zhou2023uni, luo2024bridging, yao2024node, wang2024comprehensive}. Specifically for NMRNet, We adopted and modified Uni-Mol to describe its local environment using atomic representation. Notably, NMRNet is the first application of Uni-Mol to atomic-level tasks and to extend its ability to solids in addition to gaseous and liquid states. To ensure accurate predictions across these scenarios, we collect extensive structural databases for pre-training and NMR datasets for fine-tuning, as summarized in Table \ref{tab:database_info}. 

The initial module, data preparation, is designed to extract 3D structural information from the dataset and convert it into model inputs. For liquid-state NMR, the chemical environment is based on a single molecule. In contrast, for situations requiring PBC, typically in solid-state NMR, the chemical environment of each atom is defined within the unit cell, encompassing all atoms within a specified cutoff radius. In this study, the cutoff radius is set to 6 \AA{} (the convergence check with different cutoff radius is shown in Supplementary Methods), which adequately covers the local environment for most of the atoms. The specific processing of these two representations is shown in Supplementary Fig. 1.
 
Next, NMRNet acquires general representations and knowledge from large-scale data at the pre-training stage, followed by fine-tuning that adapts the knowledge to specific tasks or domains. This paradigm significantly reduces computational costs and data requirements while improving our model's generality and specificity on target tasks. For liquid-state NMR, we directly utilized the pre-trained weights obtained from previous work by Uni-Mol on a large-scale molecular dataset\cite{zhou2023uni}. For solid-state NMR, the representation strategy based on single molecules by Uni-Mol was no longer applicable. Therefore, we adopted a reasonable representation based on cutoff radius (detailed specifics can be found in the Methods Section). Building upon this, we adjusted the pre-training strategy and performed re-pre-training using a large-scale crystal dataset. In total, we gathered over 4.8 million 3D structures from the Aflow\cite{curtarolo2012aflow}, CSD\cite{groom2016cambridge}, and Materials Project\cite{jain2013commentary} databases. The local environment of each atom is extracted for self-supervised pre-training (see Supplementary Fig. 2a). This pre-training enhances NMRNet's feature representation capabilities, allowing it to overcome limitations posed by scarce high-quality NMR data.

During the fine-tuning stage, NMRNet demonstrates versatility by supporting both simultaneous training and prediction across all NMR-active elements, as well as focused training and prediction for individual elements. This flexibility allows for tailored model optimization based on specific research needs. To comprehensively assess the NMRNet's capability, we collect as many open-source NMR datasets as possible. Liquid-state data includes nmrshiftdb2-2018\cite{jonas2019rapid, kwon2020neural}, nmrshiftdb2-2024 (organized and reported by this work) and QM9-NMR\cite{gupta2021revving, zou2023deep}. While solid-state dataset consists of ShiftML1\cite{paruzzo2018chemical}, ShiftML2\cite{cordova2022machine}, and NN-NMR\cite{lin2022machine}. For each dataset, we fine-tune NMRNet specifically for one element using pre-trained models, unless otherwise specified. We report three key metrics: Mean Absolute Error (MAE), Root Mean Square Error (RMSE), and the coefficient of determination ($R^2$). The performance of NMRNet, along with comparisons to previous studies, is summarized in Supplementary Tables 1-2. These results demonstrate NMRNet's state-of-the-art performance across the majority of datasets, highlighting its accuracy in chemical shift prediction. 

Subsequently, during the inference stage, we demonstrated NMRNet's generalizability and robustness through a series of experiments. These investigations highlighted the model's promising applications in critical areas of NMR analysis, specifically peak assignment and conformation determination. Additionally, NMRNet's structural representation exhibited the capability to elucidate the intricate relationships between local chemical environments and chemical shifts, potentially offering deeper insights into molecular structure-spectrum correlations. To make the fine-tuned NMRNet more accessible to the broader research community, we provide a user-friendly interactive web app\cite{nmrnet} that allows for quick and straightforward predictions of chemical shifts for molecules.

\begin{table}[t]
\centering
\caption{Database Information}\label{tab:database_info}

\begin{tabular}{@{}cccc@{}}
\toprule
            Database        & Data Type\footnotemark[1] & Source & Data Quantity\footnotemark[2]  \\
            \midrule
            Aflow\cite{curtarolo2012aflow}           & Structure only & Computational & 3,500,000+ \\
            CSD\cite{groom2016cambridge}             & Structure only & Experimental & 1,200,000+ \\
            Materials Project\cite{jain2013commentary}  & Structure only & Computational & 150,000+ \\
            nmrshiftdb2-2018\cite{jonas2019rapid} & Structure \&  liquid-state NMR      & Experimental & 350,000+ \\
            nmrshiftdb2-2024 & Structure \&  liquid-state NMR      & Experimental & 480,000+ \\
            QM9-NMR\cite{gupta2021revving}         & Structure \&  liquid-state NMR      & Computational & 2,300,000+ \\
            ShiftML1\cite{paruzzo2018chemical}        & Structure \&  solid-state NMR    & Computational & 250,000+ \\
            ShiftML2\cite{cordova2022machine}        & Structure \& solid-state NMR       & Computational & 470,000+ \\
            NN-NMR\cite{lin2022machine}          & Structure \& solid-state NMR      & Computational & 16,000 \\
\bottomrule
\end{tabular}
\footnotetext[1]{The databases containing only structural information were used for pre-training, while the databases with NMR information were used for fine-tuning.}
\footnotetext[2]{For structure databases, the quantity reports the number of structures. For NMR databases, it report the number of chemical shifts.}
\end{table}

\begin{figure}[h]
    \centering
    \includegraphics[width=0.9\linewidth]{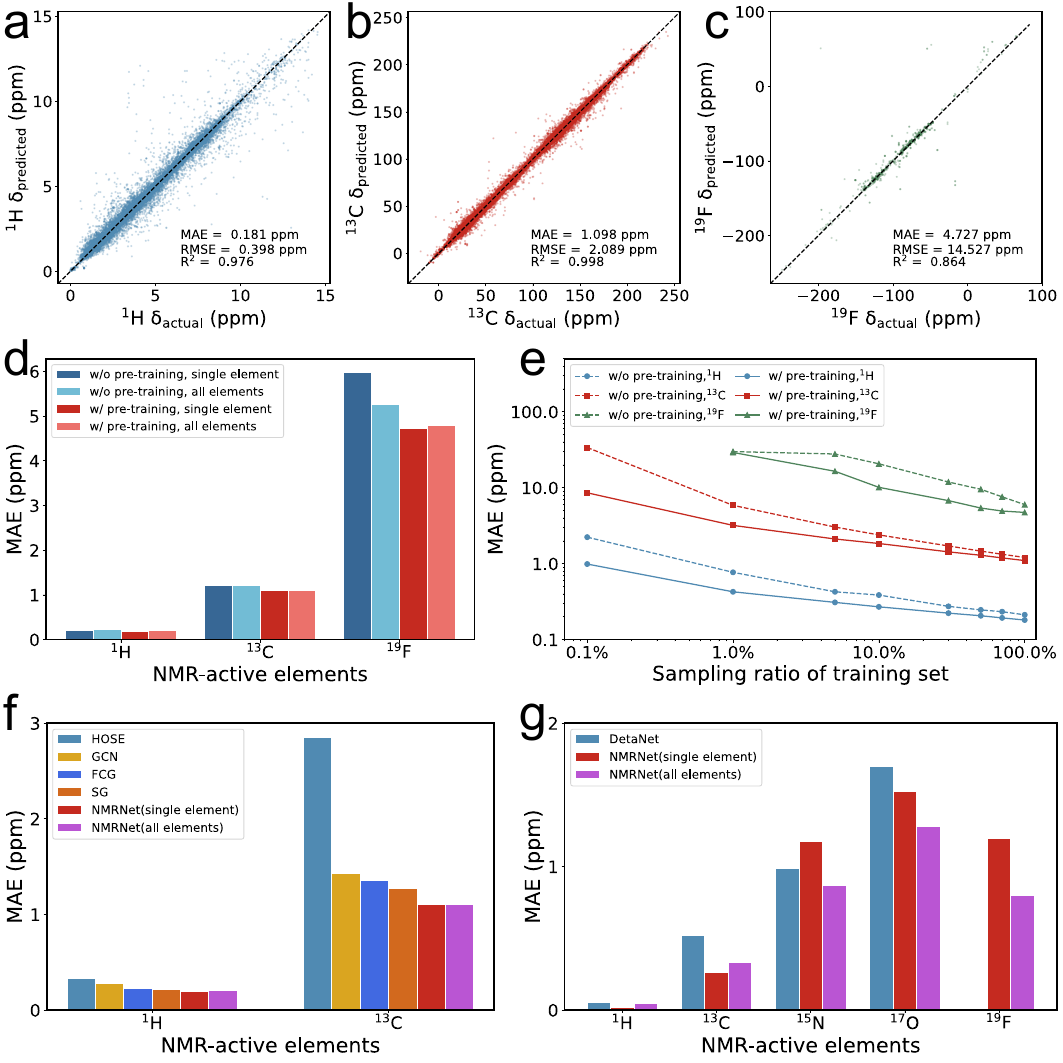}
    \caption{\textbf{Performance of NMRNet in liquid-state NMR prediction.} NMRNet's correlation scatter plots of predicted versus experimental chemical shifts for (a) $^1$H, (b) $^{13}$C, and 
  (c) $^{19}$F in the nmrshiftdb2-2024 dataset. (d) Comparison of the prediction error (MAE) for different elements in the nmrshiftdb2-2024 dataset when using or not using pre-trained weights and when predicting a single element versus all elements simultaneously. (e) Comparison of prediction error (MAE) for different elements in the nmrshiftdb2-2024 dataset using different proportions of the training set, noting that the data volume for the $^{19}$F element does not support a 0.1\% setting. To facilitate the presentation, both the horizontal and vertical axes are scaled logarithmically. Comparison of prediction error (MAE) for different elements in  (f) the nmrshiftdb2-2018 dataset and (g) the QM9NMR dataset against previous studies. Note that DetaNet has not reported results for $^{19}$F.
}
    \label{fig:lsnmr}
\end{figure}

\subsection*{Fine-tuning NMRNet with liquid-state NMR data}

As mentioned in the Introduction Section, we revisit the latest data from nmrshiftdb2, manually screen and correct erroneous entries, and ultimately create a more comprehensive and reliable dataset, nmrshiftdb2-2024 to resolve the drawback of the wide-acknowledged nmrshiftdb2-2018 dataset\cite{jonas2019rapid}. This new dataset features a greater number of atoms, a wider range of elements, and more complex structures (see Supplementary Methods and Supplementary Fig. 3a-c for details). Although the nmrshiftdb2-2024 dataset is more challenging, the final prediction error is lower compared to nmrshiftdb2-2018 (see Supplementary Fig. 3d), indicating that the model maintains excellent performance even in more complex systems. 

As illustrated in Fig. \ref{fig:lsnmr}a-c, there is a strong correlation between NMRNet's predicted results for $^{1}$H, $^{13}$C, and $^{19}$F in nmrshiftdb2-2024 and the corresponding experimental values. The prediction errors (MAE) are 0.181 ppm for $^{1}$H and 1.098 ppm for $^{13}$C, which are close to the intrinsic experimental errors previously reported \cite{jonas2019rapid}(0.09 ppm for $^{1}$H, 0.51 ppm for $^{13}$C). NMRNet also shows a reliable predictive capability for $^{11}$B and $^{17}$O, with $R^2$ values greater than 0.80, despite these elements being represented by fewer than 230 molecules in the training set (see Supplementary Table 1). To demonstrate the effectiveness of pre-training, we compare the fine-tuned NMRNet results on nmrshiftdb2-2024 using models with and without pre-training. As shown in Fig. \ref{fig:lsnmr}d, prediction errors significantly decrease when pre-training is included. Additionally, the pre-trained NMRNet shows excellent accuracy in predicting both all elements simultaneously and predicting single element during fine-tuning. As shown in Fig. \ref{fig:lsnmr}e, the model's performance improves as the training set size increases, regardless of whether pre-training is used. Still, the model with pre-training outperforms the one without, especially when the training set is small. Even when the full training set is used (sampling ratio = 100\%), the model with pre-training outperforms with lower MAEs. This emphasizes the benefit of leveraging large-scale self-supervised pre-training when fine-tuning data is limited.

Moreover, to highlight the advantages of our strategy, we compare it with previous studies\cite{bremser1978hose, jonas2019rapid, kwon2020neural, han2022scalable, zou2023deep} on the nmrshiftdb2-2018 (see Fig. \ref{fig:lsnmr}f) and QM9-NMR datasets(see Fig. \ref{fig:lsnmr}g). NMRNet shows a further improvement in the prediction accuracy of $^1$H and $^{13}$C elements compared to the previous best methods on the nmrshiftdb2-2018 dataset. Since chemical shifts in QM9-NMR can be calculated by subtracting chemical shielding from reference values, we set shielding values as training targets. The training/test splits for QM9-NMR follow the settings established by DetaNet\cite{zou2023deep}. NMRNet consistently outperforms across all six environments (1 gas + 5 solvents), with significant improvements in MAE for $^1$H (from 0.054 ppm to 0.020 ppm) and $^{13}$C (from 0.520 ppm to 0.262 ppm). This highlights NMRNet's strong ability to accurately model diverse solvent environments.

\begin{figure}
    \centering
    \includegraphics[width=0.9\linewidth]{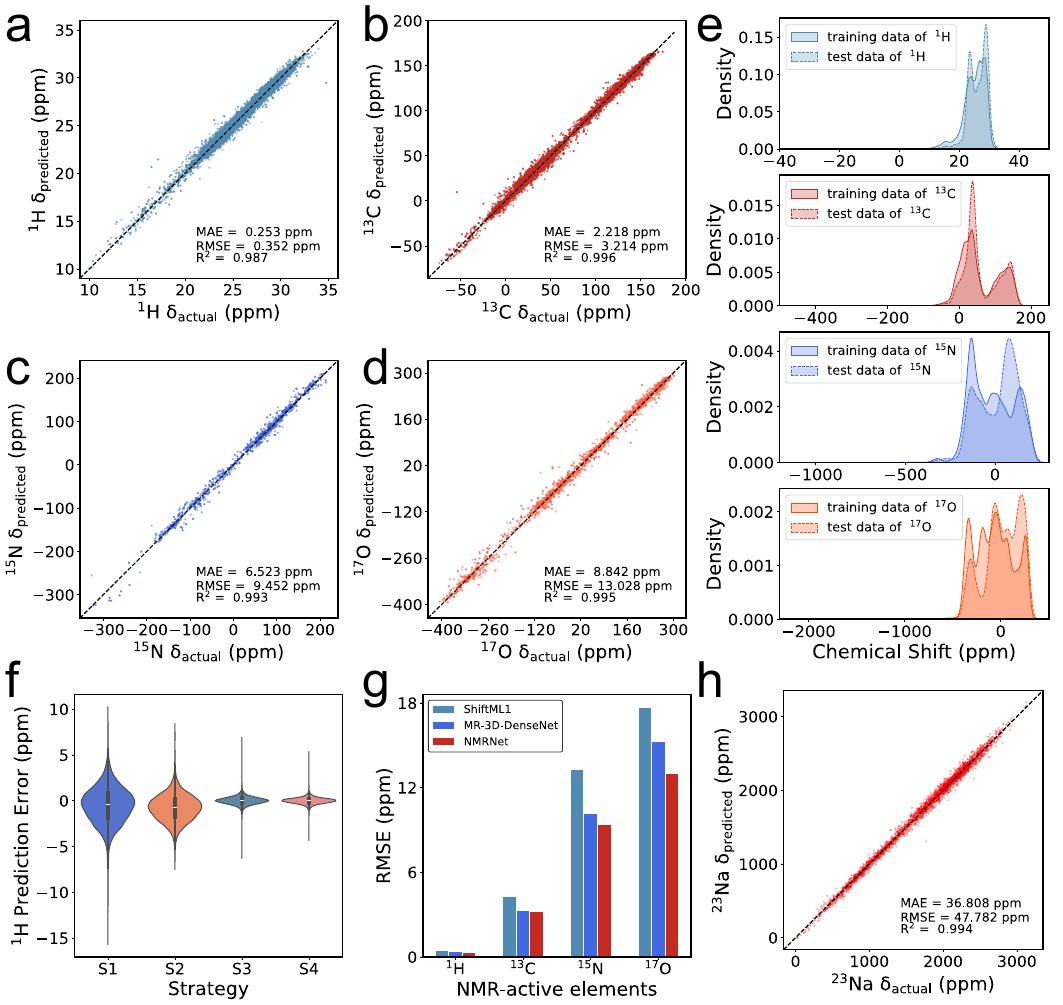}
    \caption{\textbf{Performance of NMRNet in solid-state NMR prediction.} NMRNet's correlation scatter plots of predicted versus DFT-calculated chemical shifts (chemical shieldings) for (a) $^{1}$H, (b) $^{13}$C, (c) $^{15}$N, and (d) $^{17}$O in the ShiftML1 dataset. (e) Distribution of chemical shifts for four elements in the ShiftML1 dataset. (f) The impact of four strategies on the prediction error (MAE) for $^{1}$H in the ShiftML1 dataset using NMRNet. S1-S3 utilized pre-trained weights on molecular dataset, differing in their use of the unit cell with intra-cell distance matrix, the unit cell with global distance matrix, and cutoff radius = 6 \AA{} as the local environment for a single atom, respectively. S4 modifies the pre-training in S3 to pre-training with the cutoff format on a large-scale crystal database. (g) Comparison of the prediction error (RMSE) for different elements in the ShiftML1 dataset using NMRNet with previous studies. (h) NMRNet's correlation scatter plot of predicted versus DFT-calculated chemical shifts (chemical shieldings) for $^{23}$Na in the NN-NMR dataset.
}
    \label{fig:ssnmr}
\end{figure}

\subsection*{Fine-tuning NMRNet with solid-state NMR data}

To enhance the model's understanding of solid-state NMR, we tested four different approaches. The first strategy (S1, Uni-Mol's original representation strategy) utilizes pre-trained weights from previous work\cite{zhou2023uni}, and incorporates the unit cell with an intra-cell distance matrix. The second strategy (S2) employs the same pre-trained weights as S1 but uses the unit cell with a global distance matrix. In the third strategy (S3), we maintain the pre-trained weight setup but introduce a cutoff radius of 6 \AA{} to define the local environment. Finally, the fourth strategy (S4) was similar to S3, but the pre-training weights are adapted to the cutoff format using a large-scale crystal database.

To validate the performance differences resulting from the four strategies, we fine-tuned the model on the ShiftML1 dataset using each strategy. The error distributions of the model predictions under these four strategies are shown in Fig.\ref{fig:ssnmr}f and Supplementary Fig. 4. For inputs involving the unit cell, using the global distance matrix (S2) provides more accurate descriptions than the intra-cell distance matrix (S1). Moreover, the S3 of setting the cutoff radius offers a more comprehensive depiction of the chemical environment and is expected to further improve with crystal pre-training. Therefore, we adopt S4, which resulted in a high correlation between its predicted results and DFT calculations for the four elements $^{1}$H, $^{13}$C, $^{15}$F, and $^{17}$O in the ShiftML1 dataset (see Fig. \ref{fig:ssnmr}a-d). Similar strong performance was observed on the ShiftML2 dataset (see Supplementary Table 2). It's important to note that the training set is sampled using farthest point sampling (FPS), while the test set is randomly selected. This sampling method creates an extrapolated test set that includes complex chemical environments not found in the training set\cite{paruzzo2018chemical, liu2019multiresolution, cordova2022machine}. 
The rigor of this approach is illustrated by the distribution of chemical shifts in the training and test sets (see Fig. \ref{fig:ssnmr}e and Supplementary Fig. 5 for ShiftML1 and ShiftML2, respectively). 
Moreover, Fig. \ref{fig:ssnmr}g clearly illustrates the significant performance improvements of NMRNet using Strategy 4 over previous methods\cite{paruzzo2018chemical, liu2019multiresolution}, especially for $^{15}$N and $^{17}$O. 

In the previous work, Cheng et al. calculated the dynamic chemical shifts of P2-type cathode materials and then used SOAP to extract local structural descriptors, which were subsequently used by a neural network to predict chemical shifts\cite{lin2022machine}. NMRNet successfully reduces the prediction error (RMSE) of the chemical shifts of $^{23}$Na for P2-type Na$_{2/3}$(Mg$_{1/3}$Mn$_{2/3}$)O$_2$ from 125 ppm in previous work to 48 ppm (see Fig. \ref{fig:ssnmr}h), demonstrating a significant performance improvement brought by our method.

\subsection*{Applications of the fine-tuned NMRNet}

\begin{figure}
    \centering
    \includegraphics[width=0.9\linewidth]{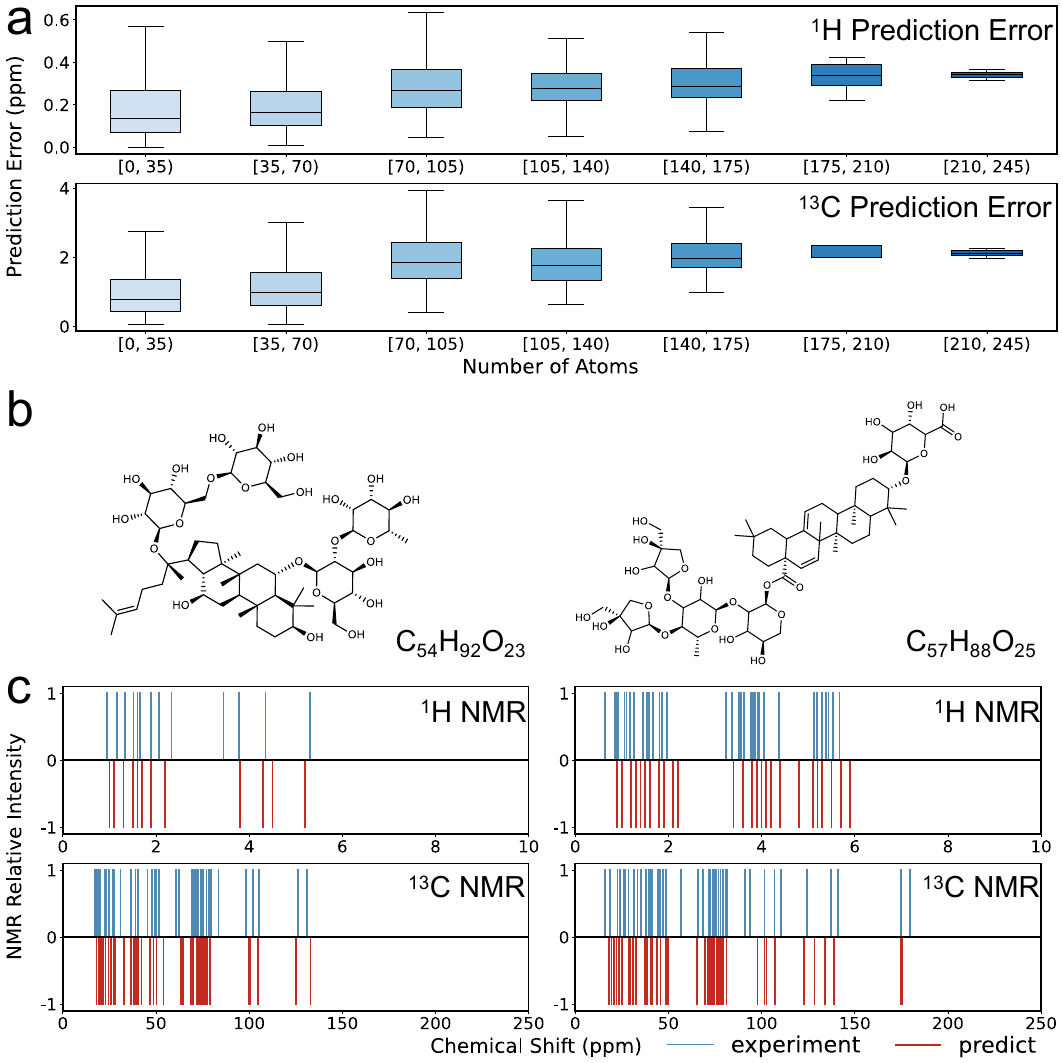}
    \caption{\textbf{Generalization test results of NMRNet.} (a) Prediction errors on the test set from nmrshiftdb2-2018 (molecules with atom counts $\leq 64$) and an additional test set from nmrshiftdb2-2024 (molecules with atom counts $\geq 70$) with respect to the distribution of the number of atoms. The top plot represents the errors for $^{1}$H, while the bottom plot represents the errors for $^{13}$C. (b) Two molecules with a relatively high number of atoms were selected to demonstrate the performance of NMRNet. Left: Yesanchinoside E ($\mathrm{C}_{54}\mathrm{H}_{92}\mathrm{O}_{23}$, nmrshiftdb2 id: 20173355), Right: Chiococcasaponin I ($\mathrm{C}_{57}\mathrm{H}_{88}\mathrm{O}_{25}$, nmrshiftdb2 id: 20253108). (c) Comparison of predicted chemical shifts with experimental chemical shifts for the two molecules in (b). The top plot shows $^1\mathrm{H}$ NMR predictions, and the bottom plot shows $^{13}\mathrm{C}$ NMR predictions. Blue vertical lines mean experimental chemical shifts, while red vertical lines are predicted chemical shifts. Peak intensities are all normalized to 1.}
    \label{fig:infer_predict}
\end{figure}

Once NMRNet is fine-tuned, it is ready to be applied in various scenarios without the need for retraining. A comprehensive exploration of fine-tuned NMRNet's robustness and generality is tested and reported in the following sections.

\subsubsection*{Generalization of the NMRNet}

Generalization is a key performance criterion for evaluating data-driven methods. To test the generalization capability (the ability to handle larger and more complex molecules) of fine-tuned NMRNet, we examine NMRNet on molecules that are not encountered during the fine-tuning phase.

Our first experiment to test the generality of NMRNet (all elements, fine-tuned on the nmrshiftdb2-2024 dataset) focuses on the task of chemical shift prediction for five nerve agents, which is crucial for evaluating the proposed procedures and preparing for future threats from new variants.\cite{jeong2022precisely} The experimental results demonstrate that the prediction accuracy of NMRNet is comparable to DFT methods (see Supplementary Fig. 8), without any preliminary knowledge or computational requirements for the nerve agents. 

Since the five nerve agent molecules have relatively simple structures, we next tested the NMRNet (all elements), fine-tuned on the nmrshiftdb2-2018 dataset (the maximum number of atoms is 64), on a more challenging scenario. We select all molecules containing more than 70 atoms from the nmrshiftdb2-2024 dataset to serve as a more challenge test set. As shown in Fig. \ref{fig:infer_predict}a, the prediction error shows no significant changes with the increase of the number of the atoms. Specifically, as shown in Supplementary Fig. 9, when using all molecules with more than 70 atoms as the test set, the $R^2$ values between the predicted and experimental results are 0.955 for $^1$H and 0.996 for $^{13}$C. Although there is a slight decrease in predictive accuracy compared to the nmrshiftdb2-2018 test set (0.971 for $^1$H and 0.998 for $^{13}$C), the predictions still show a very high correlation with the experimental values, indicating minimal overfitting. When all molecules containing over 100 atoms are used as the test set, the $R^2$ values are 0.954 for $^1$H and 0.997 for $^{13}$C, further demonstrating NMRNet's strong generalization and extrapolation robustness. To further challenge NMRNet and assess its performance in extreme cases, we extract two molecules with more than 150 atoms (see  Fig. \ref{fig:infer_predict}b) and predict their $^1$H and $^{13}$C NMR spectra (see Fig. \ref{fig:infer_predict}c). Despite the complexity of these molecules, NMRNet's performance remains satisfactory.

\subsubsection*{Peak assignment}

Peak assignment requires matching NMR signals and the atomic environment. Specifically for this task, we develop a module where the experimental chemical shifts can be assigned to the atoms corresponding to the predicted values. The results for the five nerve agent molecules are summarized in Supplementary Table 3-12. For the assignment of the $^{13}$C NMR, NMRNet achieved an accuracy of 91\%. Due to the smaller values and closer differences in experimental chemical shifts between different atoms, this task is more challenging in the $^1$H NMR. The results show that NMRNet only achieved an accuracy of 72\% for the assignment of the $^1$H NMR. 

Current NMR models have made strides in reducing chemical shift prediction errors, but challenges persist in peak assignment, which often relies on 2D NMR techniques for accuracy in experimental settings. Due to a lack of data that describes splitting and coupling between atoms, especially for protons, this information has not been effectively captured by deep learning models. Additionally, identifying chemically equivalent atoms in a molecule remains difficult, particularly when relying solely on 2D topology (i.e. graph neural network for molecular representation). While NMRNet's 3D coordinate-based approach provides a better representation of atomic environments, it still struggles with accurately distinguishing chemically equivalent atoms. These limitations hinder NMRNet's ability in peak assignment tasks but also highlight the direction for future development of NMR-related and spectrometry-related deep learning models.

\subsubsection*{Configuration determination}

In real-world cases, one of the most challenging aspects of NMR spectrum interpretation is identifying configuration and stereochemistry, particularly because isomers often have very similar chemical environments. Incorrect identification of molecular structure can lead to dramatically altered physical and chemical properties, potentially turning a therapeutic drug into a toxic compound. Moreover, an incorrect assignment of a natural product's structure can result in years of wasted effort, as researchers may pursue the synthesis of an entirely incorrect molecular structure. Traditional methods, such as the empirical Karplus equation which relies on dihedral angle assumption, address this challenge. Computational-based methods, such as DP4\cite{smith2010assigning}, show promise in distinguishing chiral isomers. Furthermore, using DFT calculated chemical environment descriptors, the model by Gao et al.\cite{gao2020general} can distinguish the chemical shifts of hyacinthacine isomers, though retraining with data from two of the isomers is required. Wu et al.\cite{wu2023elucidating} recently demonstrated the potential of machine learning models to elucidate complex organic compound structures using $^{13}$C NMR chemical shifts. In a subsequent study, Wang et al.\cite{ai2024very} recently reported the isomerism determination capability of deep learning methods based on GFN2-xTB (semi-empirical method) and GCN, similarly utilizing $^{13}$C NMR chemical shifts. However, there are limited reports of successful chiral isomer identification without the aid of computational chemistry. By incorporating 3D molecular representations, we propose that NMRNet can overcome the limitations of 1D and 2D molecular information and accurately solve this complex task without requiring extensive prior chemical knowledge.

\begin{figure}
    \centering
    \includegraphics[width=0.9\linewidth]{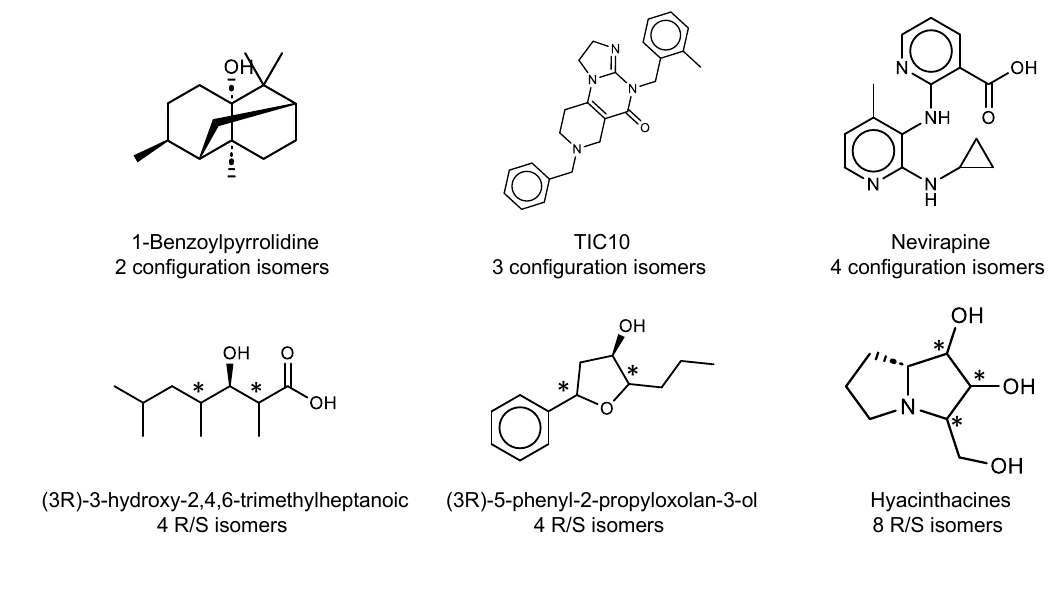}
    \caption{\textbf{Six examples used in configuration determination.} The top three are for the structure revision task, and the bottom three are for the chiral isomer identification task. The asterisk denotes a chiral center.}
    \label{fig:infer_conf}
\end{figure}

After thoroughly reviewing reported isomer cases in related NMR models and cross-referencing the original experimental papers that detailed the synthesis and characteristics of these molecules, we select six cases (see Fig. \ref{fig:infer_conf}) with the most comprehensive data and detailed reported results from previous studies. This selection ensures a fair evaluation of NMRNet's performance. The structural determination is conducted by comparing the experimental chemical shifts with the predicted chemical shifts of various candidate structures. Then, the corresponding configuration is then identified by determining the structure with the lowest Root Mean Square Deviation (RMSD) between the predicted and experimental values.

In the structure revision task, we successfully revise the structures of 1-Benzoylpyrrolidine, TIC10, and Nevirapine among pairs of close isomers using only $^{13}$C NMR data (see Figure~\ref{fig:infer_conf} top and Supplementary Table 13-15). Moving to the more challenging task of chiral isomer identification, specifically R/S stereochemistry determination (see Figure~\ref{fig:infer_conf} bottom and Supplementary Table 16-22), NMRNet identifies 3 out of 4 isomers of (3R)-3-hydroxy-2,4,6-trimethylheptanoic using $^{13}$C NMR data alone. Another example is (3R)-5-phenyl-2-propyloxolan-3-ol, where all four possible isomers are successfully identified when both $^1$H and $^{13}$C NMR data are considered. However, when only $^1$H or $^{13}$C data is used, only two isomers are correctly assigned, highlighting the importance of combining both $^1$H and $^{13}$C NMR data, which aligns with chemists' experimental approaches. In the case of Hyacinthacines, a compound with eight possible isomers (seven synthesized and one yet to be synthesized), five of the eight isomers are correctly assigned using either $^{13}$C predicted chemical shifts, even when the atoms in the diastereomers share very similar chemical environments. When both $^1$H and $^{13}$C NMR results are combined, the correct isomer's RMSD decreases, demonstrating that integrating multiple NMR datasets facilitates more accurate structure assignment. This approach underscores the critical role of comprehensive NMR analysis in stereochemistry determination, offering a robust pathway for accurate molecular identification.

Moving forward, it will be crucial to collect and standardize more NMR data on chiral isomers to enhance deep learning methods, with the aim of training models that are more sensitive to these subtle differences. Additionally, combining deep learning identification methods with other characterization techniques, such as circular dichroism, could further streamline this complex task.

\begin{figure}
    \centering
    \includegraphics[width=0.9\linewidth]{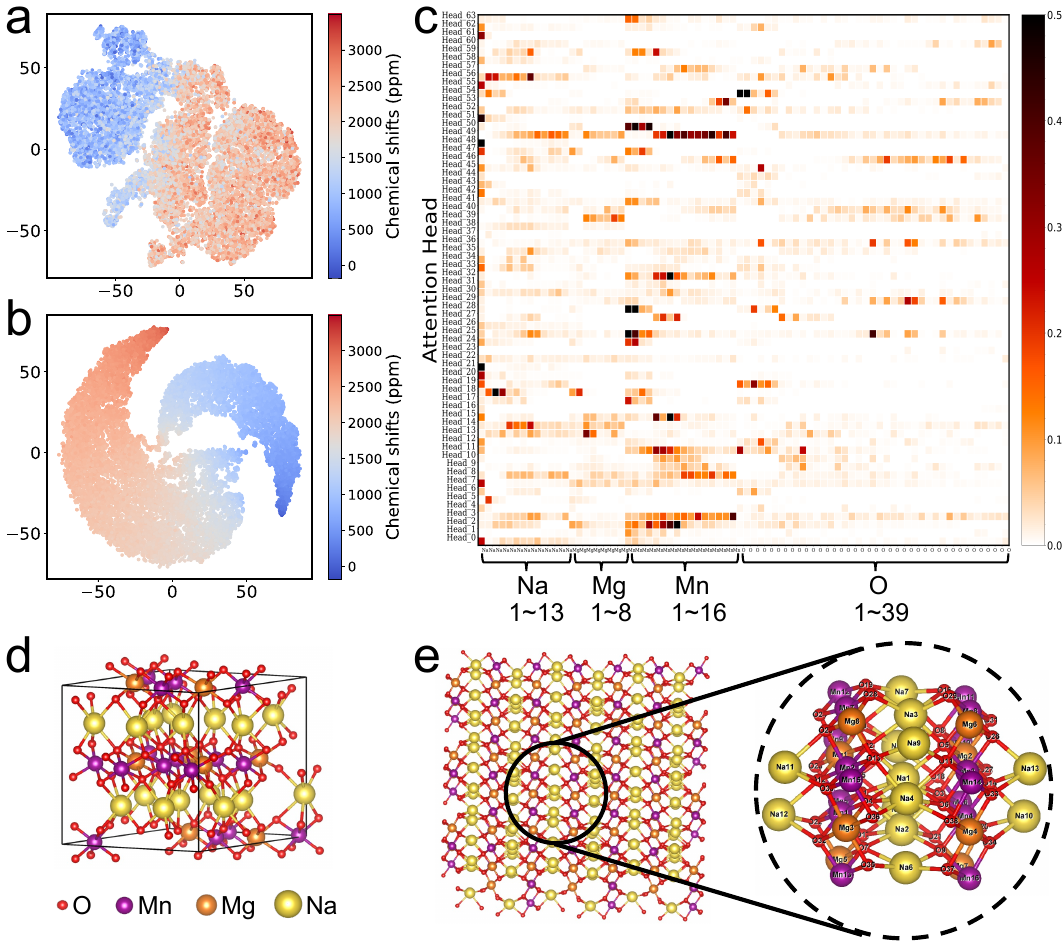}
    \caption{\textbf{Structural representations by NMRNet.} Local structural representations of and their relationship with chemical shifts for all Na$^+$ in P2-type Na$_{2/3}$(Mg$_{1/3}$Mn$_{2/3}$)O$_2$ using t-SNE for the (a) pre-trained NMRNet and (b) fine-tuned NMRNet. (c) Extract the interaction information between each central atom (represented as Na1) and its local environment (Na$_{13}$Mg$_8$Mn$_{16}$O$_{39}$) from the results of the 64-head attention mechanism of the Transformer, each head's results are represented as a separate row, and these results are then concatenated together. Identical elements are arranged in ascending order based on their distances from the central atom. (d) A unit cell of Na$_{2/3}$(Mg$_{1/3}$Mn$_{2/3}$)O$_2$. (e) The local environment of Na extracted from the infinite crystal structure corresponding to the unit cell in (d).}
    \label{fig:repr}
\end{figure}

\subsubsection*{Correlation between NMR and the local environment}

Since chemical shifts reflect the atomic local environment, we validate that NMRNet can accurately represent these environments during both the pre-training and fine-tuning phases (with solid-state NMR data). This is explored by visualizing the local structural representations and their relationship with chemical shifts for all Na$^+$ in P2-type Na${2/3}$(Mg${1/3}$Mn$_{2/3}$)O$_2$ using t-distributed stochastic neighbor embedding (t-SNE). The visualization for both the pre-trained and fine-tuned NMRNet models (see Fig. \ref{fig:repr}a-b) clearly shows that the structural representations learned during pre-training could preliminarily distinguish structures with different chemical shifts. After fine-tuning, the distinction becomes much more pronounced, indicating that the refined representations effectively reflects the distribution of local environments.

To delve deeper into the interactions within the structure, particularly how the central atom in the local environment is influenced by surrounding atoms, we visualized the multi-head attention mechanism of the Transformer model. In this study, a 64-head attention mechanism was employed, focusing on a local environment in P2-type Na${2/3}$(Mg${1/3}$Mn${2/3}$)O$2$ as a case study (see Fig. \ref{fig:repr}d-e). We extract the interaction details of each attention head concerning the central atom, represented here as Na1, and its entire local environment (Na${13}$Mg$8$Mn${16}$O${39}$), and then concatenate these interactions (see Fig. \ref{fig:repr}c, where identical elements are arranged in ascending order based on their distances from the central atom). The darker color in the visualization indicates stronger correlations between the central atom and its local environment. Some attention heads show strong interactions between all Mn$^{4+}$ ions in the local environment and the central Na$^+$ ion is captured, unveiling that Mn$^{4+}$ can affect the electrostatic potential and electronic environment around Na$^+$. Additionally, other attention heads reveal strong interactions between the central Na$^+$ and certain Mg$^{2+}$ or O$^{2-}$ ions, which further highlights how different cations and anions influence the local chemical environment.

This mechanistic study provides a tool for analyzing the intricate interactions within the local environment, bypassing the need for prior chemical knowledge. By revealing these atomic-level interactions, NMRNet enhances the interpretation of solid-state NMR spectral data, particularly in understanding how different local environments contribute to observed chemical shifts, thus aiding in the determination of structure-property relationships in complex materials.

\subsubsection*{Conclusion}

In summary, this work lays a foundational stone in the evolution of deep learning-based prediction of NMR chemical shifts, proposing and validating the potential of a unified framework for this research direction. We introduced NMRNet, pioneering the application of SE(3) Transformer-based representation learning models and a paradigm that integrates pre-training with fine-tuning for NMR prediction. This innovative approach enables rapid and precise chemical shift predictions across diverse scenarios, achieving state-of-the-art performance on nearly all datasets. Remarkably, NMRNet demonstrates significant promise in complex tasks such as structural assignment and stereochemistry determination. Furthermore, we have established a comprehensive benchmark for NMR chemical shift prediction, designed to impartially and rigorously evaluate various methods across different datasets, thereby fostering the continued advancement of this field. Moreover, our pre-trained model has established an efficient representation of the local environment of crystals, which can be directly applied to various downstream tasks related to crystalline materials. 

Future work should focus on incorporating experimental factors like solvents and temperature into the model design, as well as considering atomic coupling effects. These improvements would enhance the model's predictive accuracy and expand its applicability. Although the path ahead is challenging, the door to AI-assisted spectral interpretation has been opened. NMRNet brings immense hope for advancing spectral analysis and structural elucidation, while also providing a deeper understanding of the complexities involved. This work represents a significant step toward realizing the full potential of AI in revolutionizing the fields of spectroscopy and structural chemistry.

\section*{Methods}\label{sec:methods}
%\label{sec:methods} !!!Removed by FJ 20240811

\subsection*{Data preparation process}

For both the pre-training and fine-tuning stages, extraction of the list of atom types and their corresponding 3D coordinates from the original dataset is essential. This process can be facilitated by RDKit\cite{RDKit}, Atomic Simulation Environment (ASE)\cite{larsen2017atomic}, or Python Materials Genomics (pymatgen)\cite{ong2013python} to extract 3D information from various types of molecular sources.

The previous nmrshiftdb2-2018 dataset utilizes data published by Chio et al.\cite{kwon2020neural}, which retains the original training/test set division by Kuhn et al\cite{jonas2019rapid}. To build nmrshiftdb2-2024 dataset (introduced by this work), we extract all valid data from nmrshiftdb2\cite{kuhn2015facilitating}. 
Then, we use EmbedMolecule and MMFFOptimizeMolecule options of Rdkit to generate 3D conformations of the molecules. 
For duplicate NMR data records, the median of multiple experiments was used. 
Additionally, InChIKeys were generated using RDKit as the unique ID for each molecule, to ensure that there are no duplicate molecules in the training and test sets. 
The dataset underwent screening and cleaning to remove recording errors (detailed processing steps are described in the Supplementary Methods). While the QM9NMR dataset follows the latest research settings, we could not locate the specific division of NMR data according to DetaNet. Consequently, we reconstructed the original training/test set division based on data reported in its Supplementary Information. For pre-training and solid-state NMR, we utilized the 3D structures from their original datasets, employing ASE or pymatgen to read and analyze the local environment of each atom. The ShiftML1\cite{paruzzo2018chemical} and ShiftML2\cite{cordova2022machine} datasets adhere to their original divisions, while the NN-NMR\cite{lin2022machine} dataset follows the setup previously established by our research group. 

After extracting the 3D information, it is further converted into a list of atomic types and a matrix representing the pairwise distances between atoms, serving as input for the model.

\subsection*{NMRNet framework}

\subsubsection*{Pre-training}

Currently, several studies have attempted to mix data from different scales and chemical systems for pre-training, thereby constructing a unified atomic model across scales\cite{abramson2024accurate, zhang2023dpa}, however, the problem of unified modeling and training has not yet been fully resolved. In this work, our pre-training database comprises Aflow, CSD, and Materials Project. The CSD primarily consists of organic crystals, while the other two databases mostly include inorganic crystals. Given the significant differences in chemical environments between organic and inorganic crystals, and the varying structural sources of the different datasets, we found that mixed pre-training does not yield better results than pre-training on separated chemical systems (see Supplementary Fig. 13). Therefore, we conducted separate pre-training for organic and inorganic crystals. Detailed hyperparameter configurations are provided in Supplementary Table 23.

In the crystal pre-training phase, we employed two tasks similar to the mask token tasks in Bert\cite{devlin2018bert} and Uni-Mol\cite{zhou2023uni}: masked atom type prediction and 3D coordinate reconstruction. For each atom's local environment, we replaced the center atom and 15\% of randomly selected atoms with [mask] tokens and added ±1 Å noise to their coordinates. A key distinction is that the model only needs to be dedicated to extracting the local environment of the center atom during this phase. Consequently, the additional heads only need to reconstruct the center atom, rather than addressing all masked atoms.

A significant challenge we addressed was the substantial imbalance in the number of local structures centered around different elements within the dataset. This imbalance made it difficult for the model to adequately focus on the local structures of less common elements. For instance, in the processed CSD dataset containing 92 elements, local structures centered around H or C elements alone accounted for over 84\% of the total. To mitigate this issue, we implemented a log-weighted resampling approach to balance the distribution of different local structures within the dataset. The results of this resampling, shown in Supplementary Fig. 2b, ensure that the model can effectively learn the chemical environments of various local structures, including those of less common elements.

\begin{equation}
	p_i = \frac{\log(x_i)}{\sum_{j=1}^{n} \log(x_j)}
\end{equation}

where  $p_i$ is the sampling probability of the $i$-th element in each training iteration after resampling, $x_i$ represents the number of samples in the original dataset that belong to the local structure centered around the $i$-th element, and $n$ is the total number of elements in the dataset.

\subsubsection*{Fine-tuning}

In the fine-tuning phase, the model is trained to predict the chemical shift of each atom based on the representation of its local environment. For the prediction of liquid-state NMR, the chemical environment is described by a single molecule. 
Specifically, the model is provided with a list of atom types in the molecule and a distance matrix of pairwise atomic distances to obtain the representation of each atom's local environment. For solid-state NMR prediction, we comparatively evaluated four distinct strategies to assess the model's capability in representing the local crystal environment.

\begin{equation}
L = \left[\ell_1, \ell_2, \ell_3\right] \in \mathbb{R}^{3 \times 3}
\end{equation}

\begin{equation}
A_{\text{unitcell}} = \left[a_1, a_2, \ldots, a_n\right] \in \mathbb{R}^{3 \times n}
\end{equation}

\begin{equation}
A_{\text{infinite}} = \left\{a_i^{\prime} \mid a_i + k_1\ell_1 + k_2\ell_2 + k_3\ell_3, k_1, k_2, k_3 \in \mathbb{Z}, i \in \mathbb{Z}, 1 \leq i \leq n\right\}
\end{equation}

\begin{equation}
A_{\text{i\_cutoff}} = \left\{a_j^{\prime\prime} = a_j^{\prime} \mid \left\|a_i - a_j^{\prime}\right\|_2 < r_{\text{cut}}, 1 \leq j \leq n\right\}
\end{equation}

where $n$ is the number of atoms in the unit cell, and $r_\text{cut}$ is the cutoff radius, which is set to 6 \AA{} in this work. The unit cell is defined by the lattice matrix $L$ and the set of atoms within the unit cell $A_{\text{unitcell}}$, $A_{\text{infinite}}$ represents the infinite crystal structure composed of these unit cells, and $A_{\text{i\_cutoff}}$ denotes the local environment of the \(i\)-th atom within the unit cell, which includes all atoms in the infinite crystal structure that are within the cutoff radius of this atom.

\begin{equation}
D_1 = \left\{ D[i, j] = \left\| a_i - a_j \right\|_2 \mid 1 \leq i, j \leq n \right\}
\end{equation}

\begin{equation}
D_2 = \left\{ D[i, j] = \min \left( \left\| a_i - a_j' \right\|_2 \right) \mid 1 \leq i, j \leq n \right\}
\end{equation}

\begin{equation}
D_3 = D_4 = \left\{ D[i, j] = \left\| a_i'' - a_j'' \right\|_2 \mid 1 \leq i, j \leq |A_{\text{i\_cutoff}}| \right\}
\end{equation}

where $n$ represents the number of atoms in the unit cell, $|A_{\text{i\_cutoff}}|$ is the number of atoms in the local environment of the $i$-th atom under the cutoff format, and $D_1$ to $D_4$ represent the input distance matrices for the four strategies, respectively. Strategies 1 and 2 input atoms from the unit cell, while Strategies 3 and 4 input all atoms within the cutoff radius. The first three strategies utilize the pre-trained weights obtained from previous molecular datasets, whereas Strategy 4 involves pre-training on crystal structures based on Strategy 3.

\begin{equation}
X = F_{\text{emb}}(A, D) = \left[x_{\text{CLS}}, x_1, x_2, \ldots, x_n\right]
\end{equation}

\begin{equation}
\delta_i = G_{\text{nmr}}(x_i), \quad 1 \leq i \leq n
\end{equation}

where $A$ and $D$ represent the input atom list and their distance matrix, respectively. The classification token (CLS) is a special token in the input atom list that can be used to represent the entire structure, and $n$ denotes the number of atoms in the input structure. $X$ is the set of atom representations, $x_{\text{CLS}}$ is the representation of the CLS token, $x_i$ are the individual atom representation, and $\delta_i$ is the predicted chemical shift. $F_{\text{emb}}$ is the network that obtains the local environment representation for each atom, composed of the backbone of Uni-Mol. $G_{\text{nmr}}$ is the network that utilizes these representations to predict chemical shifts, consisting of a feedforward neural network with two layers of nonlinear transformations.

\subsection*{Modeling and analysis of molecules and materials}

RDKit\cite{RDKit} is employed for reading or generating molecular SMILES, molecular objects, InChIKeys, 2D structures, and 3D conformations. ASE\cite{larsen2017atomic} and pymatgen\cite{ong2013python} are utilized for reading and processing 3D conformations of crystal structures. The Molecule Recognition application\cite{molrecognition} of Bohrium\textsuperscript{\textrm{TM}} platform is utilized for automated identification and extraction of SMILES of molecular structures from images and scientific literature. MolView\cite{bergwerf2015molview} is employed for the visualization of 3D molecular structures, while VESTA\cite{momma2011vesta} is used for the visualization of 3D crystal structures.

\section*{Data availability}

All structural datasets used for pre-training are publicly accessible. The Aflow dataset\cite{curtarolo2012aflow} is available at \url{https://aflowlib.org/}, the Materials Project dataset\cite{jain2013commentary} is accessible at \url{https://next-gen.materialsproject.org/}, and the CSD dataset\cite{groom2016cambridge} is accessible at \url{https://www.ccdc.cam.ac.uk/}. All processed NMR datasets used for fine-tuning are available at \url{https://zenodo.org/records/13317524}.

\section*{Code availability}

The NMRNet code is available at \url{https://github.com/Colin-Jay/NMRNet} under an open-source license. The trained model parameters are available at \url{https://zenodo.org/records/13317524}.

\bibliography{sn-bibliography}% common bib file
%% if required, the content of .bbl file can be included here once bbl is generated
%%\input sn-article.bbl

\end{document}